\documentclass{iaus}
\usepackage{graphics}

\title[Metallicity of M dwarfs] 
{Metallicity of M dwarfs: \\ the link to exoplanets}

\author[V. Neves, X. Bonfils \& N.C. Santos]   
{V. Neves$^{1,2}$, X. Bonfils$^2$, \and N.C. Santos$^{1,3}$}

\affiliation{$^1$Centro de Astrof{\'\i}sica, Universidade do Porto, Rua das Estrelas, 4150-762 Porto, Portugal \\ email: {\tt vasco.neves@astro.ua.pt} \\[\affilskip]
$^2$Laboratoire d'Astrophysique, Observatoire de Grenoble, BP 53, F-38041 Grenoble C{\'e}dex 9, France \\
$^3$Departamento de F{\'i}sica e Astronomia, Faculdade de Ci{\^e}ncias, Universidade do Porto, Portugal
}

\pubyear{2011}
\volume{276}  
\pagerange{1--2}
\jname{The Astrophysics of Planetary Systems: Formation, Structure, and Dynamical Evolution}
\editors{Alessandro Sozzetti, Mario G. Lattanzi \& Alan P. Boss, eds.}
\begin{document}

\maketitle

\begin{abstract}

The determination of the stellar parameters of M dwarfs is of prime importance in the fields of galactic, stellar and planetary astronomy. M stars are the least studied galactic component regarding their fundamental parameters. Yet, they are the most numerous stars in the galaxy and contribute to most of its total (baryonic) mass. In particular, we are interested in their metallicity in order to study the star-planet connection and to refine the planetary parameters. As a preliminary result we present a test of the metallicity calibrations of \cite{B05}, \cite{JA09}, and \cite{SL10} using a new sample of 17 binaries with precise V band photometry.

\end{abstract}

\firstsection 

\section*{Preliminary Results}

We tested the metallicity calibration of \cite{B05} (hereafter B05), as well as the calibrations of \cite{JA09} (hereafter JA09) and \cite{SL10} (hereafter SL10), with a sample of 17 M dwarf secondaries with a wide ($> 5$ arcsec separation) physical FGK companion. 

Following B05, three papers with different calibrations were published: JA09, SL10, and \cite{RA10} (hereafter RA10). Each work claims a calibration with a better precision than the previous ones, and in general, poor V photometry is identified as a serious limitation. In order to address the photometric limitation, only M stars with precise V photometry ( $\sigma<$ 0.04 mag) were selected. Most stars have V magnitude uncertainties of 0.01 or 0.02 mag. Note that the RA10 calibration was not tested because it requires IR indices that we do not have. This test will be done in the near future.

We found that the metallicity values of our stars (obtained from the FGK primary component) are in reasonable agreement with the [Fe/H] values obtained with all calibrations, as can be seen in Fig. 1. However, our calibrators are found to be more metal poor (on average) than both JA09 and SL10 calibrations. 

A better photometry did not improve the dispersion measured around the different calibrations. This means that precision on V photometry may not be the main limitation in the derivation of the [Fe/H] calibration.



Table 1. shows a quantitative comparison between the calibrations. We note that the rms, $RMS_{P}$ and the $R^{2}_{ap}$ values were offset-corrected. In general, the calibrations have similar offsets, $rms$, $RMS_{P}$, and correlation coefficients.

Interestingly, the calibration of B05 (1) has the lowest offset and rms. However, the correlation coefficient is a bit lower than the values of JA09 and SL10. The results are inconclusive and require further study. 


\begin{figure} 
\centering 
\resizebox{4.4cm}{!}{\includegraphics{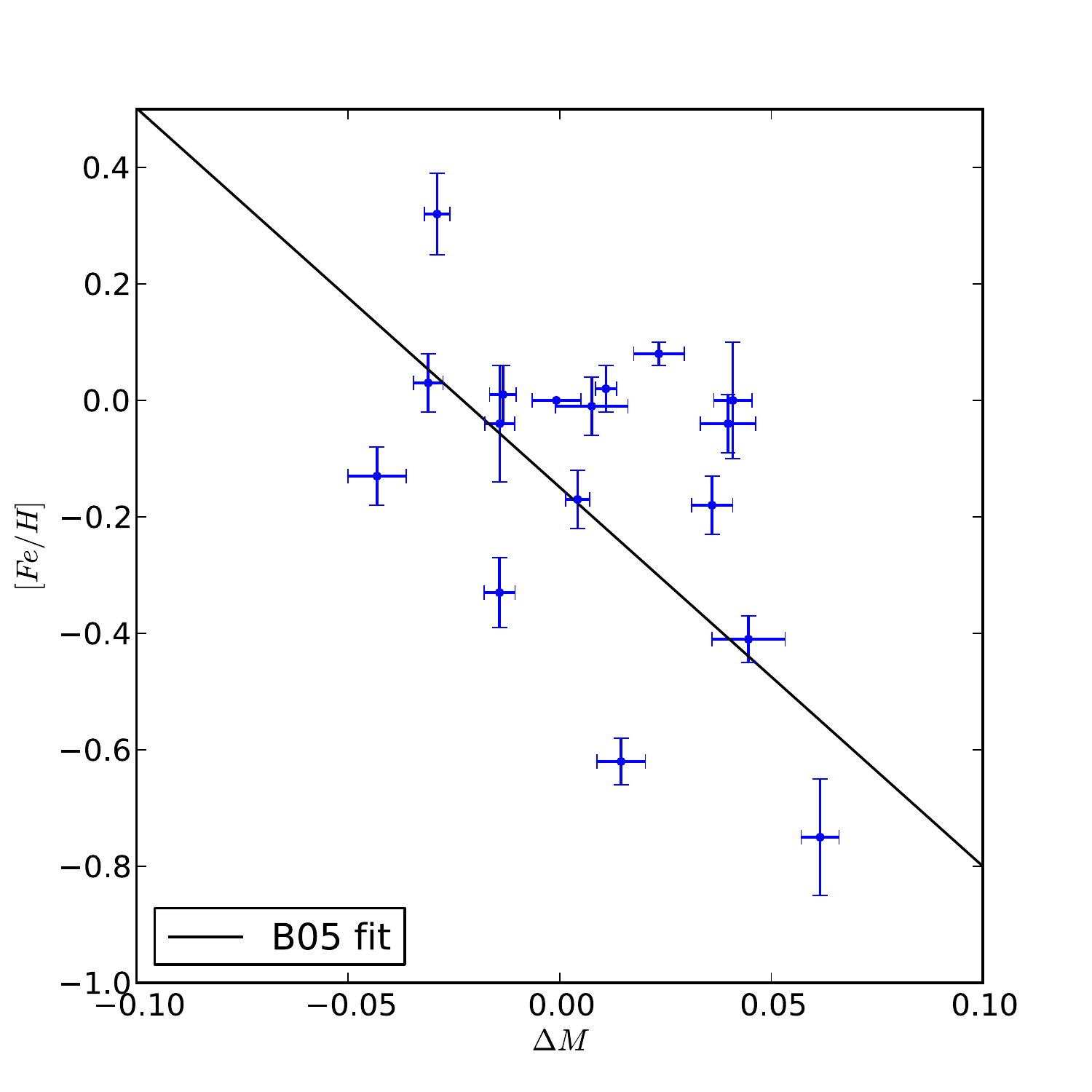}}
\resizebox{4.4cm}{!}{\includegraphics{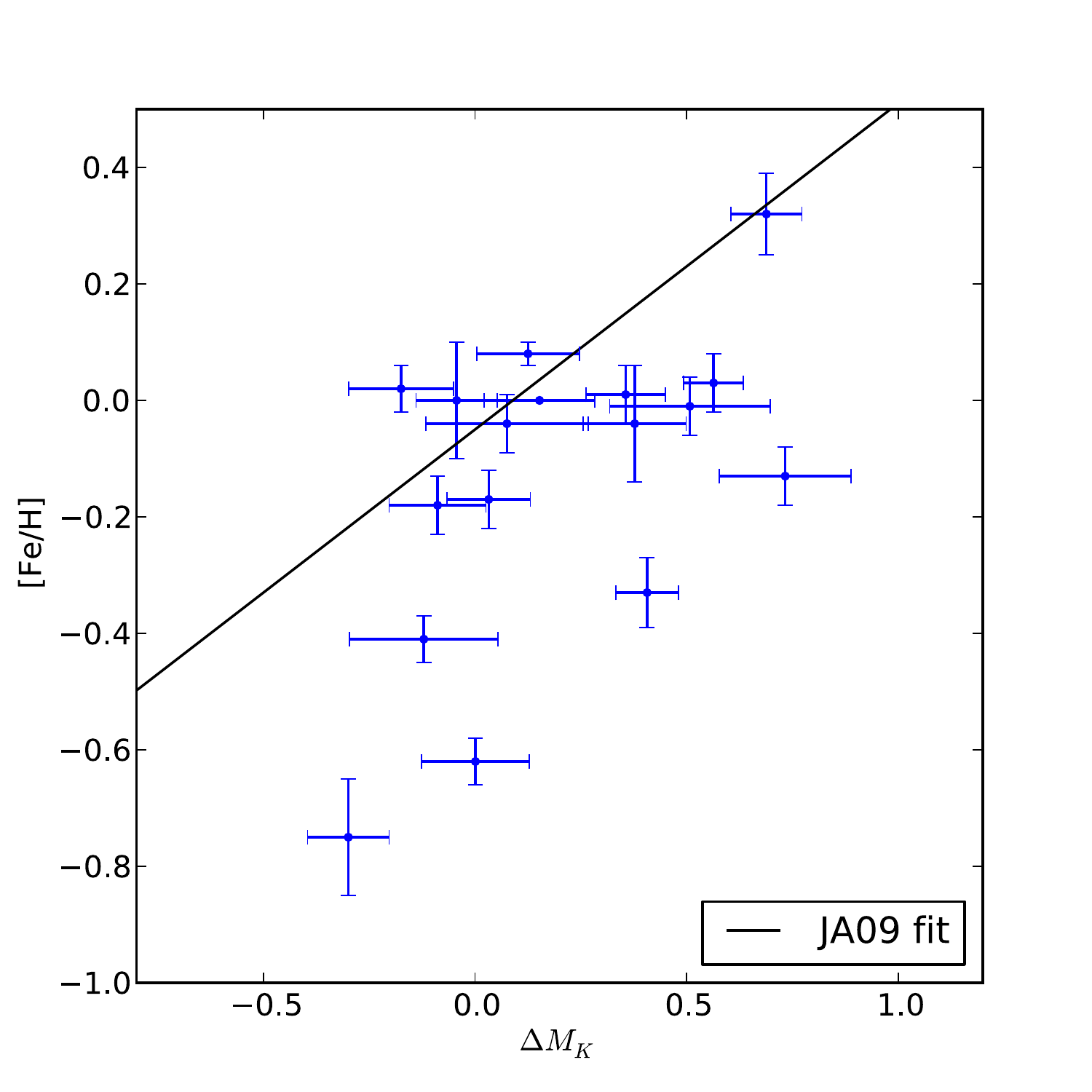} }	
\resizebox{4.4cm}{!}{\includegraphics{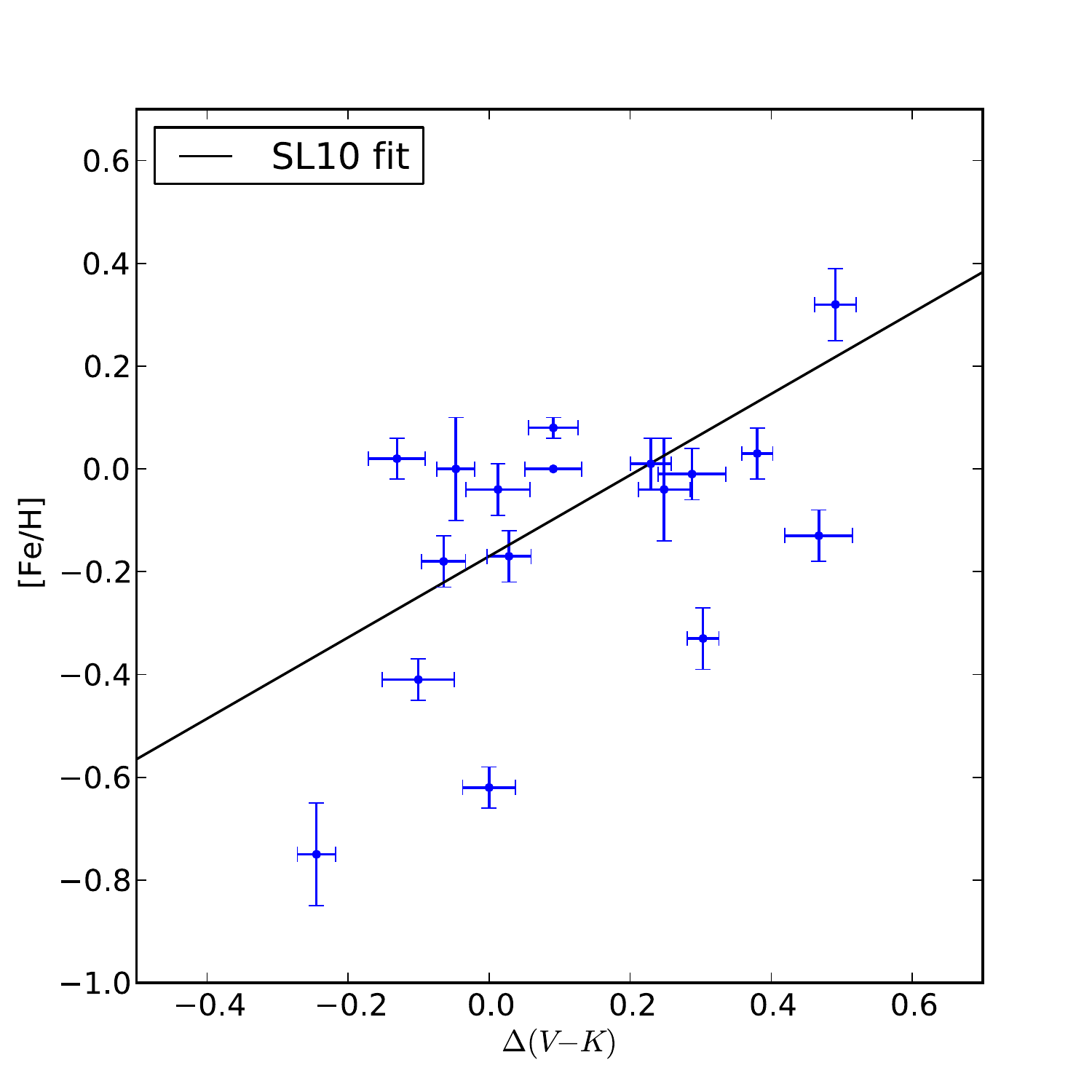} } 
\caption[]{\textit{Left panel:} Plot of metallicity versus the difference between masses calculated from the V- and the K-band Mass-Luminosity equations of \cite{D00}. The black line represents the calibration of B05 (2). \textit{Middle panel:} [Fe/H] versus the difference between the mean value of Mk of M dwarfs and the value of Mk of each star, as defined by \cite{JA09}. The black line represents the calibration of JA09. \textit{Right panel:} [Fe/H] versus the observed difference between V and K magnitudes (V-K) and the fit of the (V-K) corresponding to the horizontal distance (in the V-K, Mk plane) between the mean value of Mk of M dwarfs and the value of Mk of each star, as defined by \cite{SL10}. The black line represents the calibration of SL10.} 
\end{figure}

\begin{table}[htdp]\tiny
\caption{Comparison of the residuals offset, $rms$, residual mean square ($RMS_{P}$), and adjusted square of the multiple correlation coefficient ($R^{2}_{ap}$) of the calibrations of \cite{B05}, \cite{JA09}, and \cite{SL10} applied to our data. $RMS_{P}$ and $R^{2}_{ap}$ definitions were taken from \cite{SL10}.}
\begin{center}
\begin{tabular}{c c c c c}

\hline
Calibration Source + equation&offset&rms&RMS$_{P}$&R$^{2}_{ap}$\\
	&(dex)&(dex)&(dex)& \\
\hline
B05 (1) : $[Fe/H] = 0.196 - 1.527M_{K} + 0.091M_{K}^{2} + 1.886(V-K) - 0.142(V-K)^{2}$ & 0.05 & 0.21 & 0.06 & 0.18 \\
B05 (2) : $[Fe/H] = -0.149 - 6.508\Delta M, \Delta M = Mass_{V} - Mass_{K}$ & 0.07 & 0.23 & 0.06 & 0.17 \\
JA09 : $[Fe/H] = 0.56\Delta M_{K} - 0.05, \Delta M_{K} = MS - M_{K}$ & 0.19 & 0.22 & 0.05 & 0.27 \\
SL10 : $[Fe/H] = 0.79\Delta (V-K) - 0.17, \Delta (V-K) = (V-K)_{obs} - (V-K)_{fit}$ & 0.06 & 0.22 & 0.05 & 0.28 \\
\hline
\end{tabular}
\end{center}
\label{default}
\end{table}%

\textbf{Acknowledgements:} We acknowledge the support by the European Research Council/European Community under the FP7 through Starting Grant agreement number 239953. NCS also acknowledges the support from Funda\c{c}\~ao para a Ci\^encia e a Tecnologia (FCT) through program Ci\^encia\,2007 funded by FCT/MCTES (Portugal) and POPH/FSE (EC), and in the form of grant reference PTDC/CTE-AST/098528/2008. VN would also like to acknowledge the support from FCT in the form of the fellowship SFRH/BD/60688/2009.




\begin{thebibliography}{}

\bibitem[Delfosse et al. (2000)]{D00}
{Delfosse}, X., {Forveille}, T., {S{\'e}gransan}, D., {Beuzit}, {J.-L.}, {Udry}, S., {Perrier}, C., \& {Mayor}, M. 2000
\textit{A\&A}, 364, 217

\bibitem[Bonfils et al. (2005)]{B05}
{{Bonfils}, X., {Delfosse}, X., {Udry}, S., {Santos}, N.~C., {Forveille}, T., \& {S{\'e}gransan}, D.} 2005
\textit{A\&A}, 442, 635

\bibitem[Johnson \& Apps (2009)]{JA09}
{{Johnson}, J.~A. \& {Apps}, K.} 2009
\textit{ApJ}, 699, 933


\bibitem[Schlaufman \& Laughlin (2010)]{SL10}
{{Schlaufman}, K.~C. \& {Laughlin}, G.} 2010,
\textit{A\&A}, 519, A105

\bibitem[Rojas-Ayala et al. (2010)]{RA10}
{{Rojas-Ayala}, B., {Covey}, K.~R, {Muirhead}, P.~S. \& {Lloyd}, J.~P.} 2010,
\textit{ApJ}, 720, L113


\end{thebibliography}
\end{document}